\documentclass[aps,pra,twocolumn,groupedaddress,showpacs,superscriptaddress,scrartcl]{revtex4-1}
\usepackage{amssymb,amsmath,graphicx,epstopdf,hyperref,amsthm,dsfont,enumerate,bm}
\begin{document}
\interfootnotelinepenalty=10000

\title{Spacetime-constrained oblivious transfer}

\author{Dami\'an Pital\'ua-Garc\'ia}
\email[]{dpitalua@ulb.ac.be}
\affiliation{Laboratoire d'Information Quantique, CP 224, Universit\'e libre de Bruxelles,
Av. F. D. Roosevelt 50, 1050 Brussels, Belgium}
\affiliation{IRIF, CNRS, Unversit\'e Paris Diderot, Paris, France}

\date{\today}

\begin{abstract}
In 1-out-of-2 oblivious transfer (OT), Alice inputs numbers $x_0,x_1$, Bob inputs a bit $b$ and outputs $x_b$. Secure OT requires that Alice and Bob learn nothing about $b$ and $x_{\bar{b}}$, respectively. We define \emph{spacetime-constrained oblivious transfer} (SCOT) as OT in Minkowski spacetime in which Bob must output $x_b$ within $R_b$, where $R_0$ and $R_1$ are fixed spacelike separated spacetime regions. We show that unconditionally secure SCOT is impossible with classical protocols in Minkowski (or Galilean) spacetime, or with quantum protocols in Galilean spacetime. We describe a quantum SCOT protocol in  Minkowski spacetime, and we show it unconditionally secure.
\end{abstract}


\maketitle
\section{Introduction}
Relativistic quantum cryptography \cite{K99,K99.2,K05.2,KBMS06,CK06,C07,C09,M10.1,KMS11,K11.1,pbqc,K13,K11.2,K11.3,K12.1,HM12,K12,CK12,KTHW13,LKBHTKGWZ13,LCCLWCLLSLZZCPZCP14,KMS14,LKBHTWZ15,AK15.1,AK15.2,CCL15,FF15,AK15.3} is a recent branch of physics-based cryptography in which the combined power of quantum physics and relativity is exploited to guarantee security. It opens up possibilities for novel cryptographic tasks that do not have analogues in nonrelativistic settings \cite{KMS11,pbqc,K11.3,K13}. Furthermore, it motivates the investigation of security for tasks that were shown impossible in nonrelativistic quantum scenarios. For example, unconditionally secure protocols for bit commitment \cite{M97,LC97}, ideal coin tossing \cite{LC98}, strong coin tossing with arbitrarily small bias \cite{Kitaev02,ABDR04}, all-or nothing oblivious transfer \cite{R02,C07}, 1-out-of-2 oblivious transfer \cite{L97}, and more general two-party classical computation \cite{L97,C07,BCS12} have been shown impossible in nonrelativistic quantum settings. Although these impossibility proofs also apply to relativistic quantum settings for must of these tasks \cite{CK06,C07,BCS12}, they do not apply to bit commitment \cite{K99,K05.2}, and ideal and strong coin tossing \cite{K99.2}, for which unconditionally secure relativistic protocols have been found \cite{K99,K99.2,K05.2,K11.2,K12,LKBHTWZ15,AK15.2}.

\begin{figure}
\includegraphics{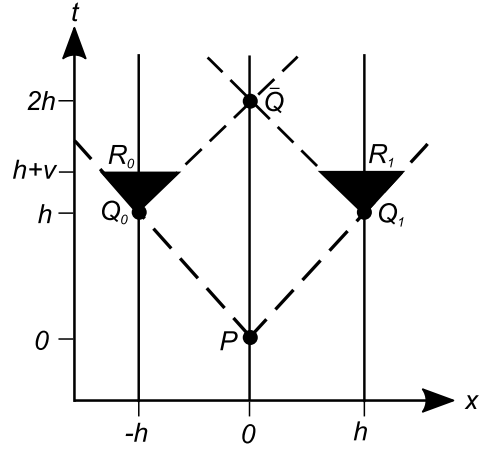}
 \caption{\label{fig}Two-dimensional spacetime diagram of the defined cryptographic task, SCOT, in the reference frame $\mathcal{F}$ in Minkowski spacetime. The vertical lines correspond to the world lines of Alice's and Bob's agents. The small circles represent the spacetime points $P=(0,0,0,0)$, $Q_0=(h,-h,0,0)$, $Q_1=(h,h,0,0)$ and $\bar{Q}=(2h,0,0,0)$. The dotted lines represent light rays, defining the future light cones of $P$, $Q_0$ and $Q_1$. The spacetime regions $R_i$, where the passwords $\bold{x}_i$ are accepted as valid by the computers $\mathcal{C}_i$, correspond to the filled regions, for $i=0,1$. $\bar{Q}$ is the spacetime point within the intersection of the future light cones of $Q_0$ and $Q_1$ with the smallest time coordinate, $t=2h$, in the frame $\mathcal{F}$.}
\end{figure}

Here we introduce a novel cryptographic task in Minkowski spacetime, which we motivate with the following problem.
We consider a particular reference frame $\mathcal{F}$ agreed a priori by Alice and Bob.
Alice has two passwords $\bold{x}_0,\bold{x}_1\in\{0,1\}^n$ at the spacetime point $P=(0,0,0,0)$, defined in $\mathcal{F}$, permitting the access to one of two computers: $\mathcal{C}_0$ or $\mathcal{C}_1$. The computer $\mathcal{C}_i$ can be accessed if the password $\bold{x}_i$ is input in the spacetime region $R_i$, for $i\in\{0,1\}$, where $R_0$ and $R_1$ are spacelike separated and within the future light cone of $P$. Here we consider that $R_i$ is the intersection of the future light cone of $Q_i=\bigl(h,-(-1)^ih,0,0\bigr)$ and the set of spacetime points $T_i^v=\bigl\{\bigl(t,x,y,z\bigr)\big\vert h\leq t\leq h+v, \bigl(x+(-1)^i h\bigr)^2+y^2+z^2\leq v^2\bigr\}$ for $i=0,1$, defined in the reference frame $\mathcal{F}$, where $v>0$ is a small parameter, $h>0$, the first entry in a four vector is temporal, and we use units in which the speed of light is $c=1$ (see Fig. \ref{fig}). The password $\bold{x}_i$ is stored securely in the computer $\mathcal{C}_i$ in the past light cone of $Q_i$. The computer $\mathcal{C}_i$ is secure from Alice and accepts the password only in the region $R_i$. Thus, the password $\bold{x}_i$ is only useful within $R_i$. Bob pays Alice at $P$ for one of the passwords. The deal is that Bob can have access to $\bold{x}_b$ in $R_b$, that Bob cannot access $\bold{x}_{\bar{b}}$ in $R_{\bar{b}}$, and that Alice should be oblivious to Bob's choice $b\in\{0,1\}$, anywhere in spacetime. We define the cryptographic task that complies with the previous conditions as \emph{spacetime-constrained oblivious transfer} (SCOT).

SCOT is a relativistic version of 1-out-of-2 oblivious transfer (OT) \cite{EGL85}. OT is an important cryptographic task that can be used to implement more sophisticated tasks \cite{K88}. As mentioned above, OT cannot be implemented with unconditional security in nonrelativistic, or relativistic, quantum cryptography  \cite{L97,CK06}. The main argument in the proof of the no-go theorem for OT is that the fact that Bob must learn Alice's message $\bold{x}_b$ with probability equal to (or very close to) unity and that, in a secure protocol, Alice does not have any information (or has very little information) about Bob's input $b$, implies that there exists a unitary operation that Bob can apply to his systems after learning $\bold{x}_0$ that transforms his systems into a state that allows him to learn $\bold{x}_{1}$ with probability equal to (or very close to) unity too \cite{L97}. Similar arguments apply in the proofs that unconditionally secure quantum bit commitment and classical deterministic two party computation are impossible \cite{M97,LC97,C07,BCS12}. 

We note that these Mayers-Lo-Chau quantum attacks \cite{M97,LC97,L97} and more general impossibility proofs \cite{C07,BCS12} do not apply to SCOT, because, due to Minkowski causality, a nonlocal unitary acting on systems in $R_0$ and $ R_1$ cannot be completed outside the intersection of the future light cones of $R_0$ and $R_1$, hence, it cannot be completed within $R_0$ or $R_1$. More generally, these no-go theorems do not apply to \emph{spacetime-constrained multi-party computation}, which is defined as an extension of multi-party computation \cite{CK06,C07,C09} in which the inputs and outputs are constrained to be input and output within fixed constrained regions of spacetime, respectively, a research problem initially outlined by Kent \cite{K11.3}.

Our results are the following. We show that unconditionally secure SCOT cannot be achieved with classical protocols in Minkowski or Galilean spacetime, or with quantum protocols in Galilean spacetime. Then, after describing the general setting of relativistic quantum cryptography, we give a quantum protocol for SCOT in Minkowski spacetime, and we show it unconditionally secure. 

Interestingly, to the best of our knowledge, SCOT is the second cryptographic task with classical inputs and outputs that is proved unconditionally secure with relativistic quantum protocols, but whose unconditional security is known impossible with classical or nonrelativistic quantum protocols, location-oblivious data transfer being the first one \cite{K11.3}.  In particular, there are relativistic classical bit commitment protocols  that are provably unconditionally secure for a single round \cite{K99,K05.2,LKBHTWZ15,AK15.2}, that is, for a duration limited by the time that light takes to travel between the labs of the participating agents.

We note that the first security proof for single-round relativistic classical bit commitment protocols was given by Kent (see Lemma 3 of Ref. \cite{K05.2}). There are conjectured unconditionally secure \cite{K99,K05.2,LKBHTWZ15} relativistic classical bit commitment protocols with multiple rounds, whose duration can be extended arbitrarily long for a fixed distance between the labs of the agents. These have been shown secure against arbitrary classical attacks \cite{K99,K05.2,LKBHTWZ15,CCL15,FF15}, but remain to be shown secure against arbitrary quantum attacks. Proving arbitrarily long relativistic, classical or quantum, bit commitment protocols unconditionally secure against arbitrary quantum attacks remains as an important open problem.

\section{SCOT is impossible in a classical or Galilean world}
\label{sec:2}

We consider arbitrary SCOT protocols taking place in a finite region of spacetime, hence, with $R_0$ and $R_1$ having a finite separation. We show that classical SCOT protocols of this form cannot be unconditionally secure, in Minkowski or Galilean spacetime. Then we argue that unconditionally secure SCOT is impossible in a Galilean quantum world, as  follows from Lo's no-go theorem \cite{L97} for nonrelativistic quantum OT.

In general we can consider that Alice and Bob have several agents who participate in the protocol at different points of spacetime.
In particular, the protocol involves communication events from an agent of Alice to an agent of Bob or vice versa at various spacetime points.
Without loss of generality we reduce the number of communication events to three, occurring at $P$, $Q_0$ and $Q_1$, respectively,
where we assume that Alice and Bob have adjacent labs of negligible extension. We consider first that spacetime is Minkowski and Alice and Bob perform a classical protocol.  We assume there exists a classical SCOT protocol that is secure against Alice. We show that this protocol must be insecure against Bob.

We give below the main physical arguments of the proof. Some mathematical details of the proof are given in Appendix \ref{appa}. Let $\mathcal{G}$, $\mathcal{G}_0$ and $\mathcal{G}_1$ be the sets of possible communication events between Alice and Bob at the spacetime points $P$, $Q_0$ and $Q_1$, respectively. Let $P(\bold{g}\vert \bold{x}_0,\bold{x}_1,b)$ be the probability that the communication event $\bold{g}=(g,g_0,g_1)\in\bar{\mathcal{G}}=\mathcal{G}\times\mathcal{G}_0\times\mathcal{G}_1$ occurs in the honest protocol, in general depending on Alice's inputs $\bold{x}_0,\bold{x}_1$ and on Bob's input $b$. Since $\bold{g}$ is classical, Alice and Bob can produce perfect copies of it, use some copies in the protocol and keep others for record. In particular, Alice's agents at $P$, $Q_0$ and $Q_1$ can send a copy of $g$, $g_0$ and $g_1$ to her agents at the other spacetime points at the speed of light, respectively. Since the protocol occurs in a finite region of spacetime, the copies of $g$, $g_0$ and $g_1$ arrive to all of Alice's agents after a finite time. Since we assume that the protocol is secure against Alice, for any fixed inputs $\bold{x}_0,\bold{x}_1$, Alice can guess Bob's bit $b$ with a bounded probability 
\begin{equation}
\label{1}
P_{\text{A}}^{\text{c}}\leq\frac{1}{2}+\delta,
\end{equation}
for some $0\leq \delta<<1$, where the labels ``c" and ``A" denote ``cheating Alice." We show in Appendix \ref{appa} that (\ref{1}) implies
\begin{equation}
\label{2}
\sum_{\bold{g}\in \bar{\mathcal{G}}}\bigl\lvert P(\bold{g}\vert \bold{x}_0,\bold{x}_1,0)-P(\bold{g}\vert \bold{x}_0,\bold{x}_1,1)\bigr\rvert
\leq 4\delta.
\end{equation}

On the other hand, following the honest protocol with any input $b\in\{0,1\}$, Bob's agents must be able to use the obtained value $\bold{g}$ and process it with some extra classical resources $E$ in order to output $\bold{x}_b$ in $R_b$ with some probability
\begin{equation}
\label{d1}
P_{\text{B}}^{\text{h}}\geq 1-\epsilon,
\end{equation}
for some $0\leq\epsilon<<1$, where the labels ``h" and ``B" denote ``honest Bob." Bob's agents can prepare and distribute two copies of $E$ before the protocol. Bob's agents can produce two perfect copies of the exchanged messages $g$, $g_0$ and $g_1$. Since the communication events $\bold{g}$ occur with probabilities $P(\bold{g}\vert \bold{x}_0, \bold{x}_1,b)$ that are very close for different values of $b\in\{0,1\}$, as given by (\ref{2}), Bob's agents can run two simultaneous protocols, one with input $b=0$ and the other one with input $b=1$ and thus output both $\bold{x}_0$ in $R_0$ and $\bold{x}_1$ in $R_1$ with probability $P_{\text{B}}^{\text{c}}$ very close to unity, satisfying 
\begin{equation}
\label{d1.1}
P_{\text{B}}^{\text{c}}\geq 1-2\epsilon-4\delta,
\end{equation}
as we show in Appendix \ref{appa}, where the labels ``c" and ``B" denote ``cheating Bob." Thus, the SCOT protocol is insecure against Bob, as claimed. This result trivially holds in Galilean spacetime because in this case Alice's agents can instantaneously communicate the values of $g$, $g_0$ and $g_1$.

Now we consider arbitrary quantum SCOT protocols in Galilean spacetime. In principle, Alice's and Bob's agents can communicate classical information instantaneously.  They can also transmit quantum states between distant locations instantaneously via quantum teleportation \cite{teleportation}. Thus, we can consider the region of three dimensional space in which the protocol takes places as two big labs adjacent to each other, one controlled by Alice and the other one by Bob. The output regions $R_0$ and $R_1$ become irrelevant in this scenario because if Bob can output $\bold{x}_b$ anywhere in his lab then he can output $\bold{x}_b$ in $R_b$. This is the scenario of nonrelativistic quantum 1-out-of-2 oblivious transfer. Therefore, Lo's no-go theorem stating that secure OT is impossible in nonrelativistic quantum cryptography applies \cite{L97}. It follows that secure SCOT is impossible in a Galilean quantum world.

\section{The setting for relativistic quantum cryptography}
\label{SIII}
We describe the setting for relativistic quantum cryptography that applies to the protocol we present here \cite{K13,K11.2,K11.3,K12.1}. Alice and Bob consist of three different agents, each controlling a secure laboratory. We denote Alice's agents by $\mathcal{A}$, $\mathcal{A}_0$ and $\mathcal{A}_1$, and Bob's agents by  $\mathcal{B}$, $\mathcal{B}_0$, and $\mathcal{B}_1$. Agents $\mathcal{A}$ and $\mathcal{B}$ have secure laboratories at space regions adjacent to the point $L=(0,0,0)$, while agents $\mathcal{A}_i$ and $\mathcal{B}_i$ have secure laboratories at space regions adjacent to the points $L_i=(-(-1)^ih,0,0)$, for $i\in\{0,1\}$. For clarity of the presentation we assume that Alice's and Bob's laboratories have negligible extension and are located exactly at $L$, $L_0$ and $L_1$ (see Fig. \ref{fig}).

We also assume that the processing of classical and quantum information at each laboratory is secure, perfect and instantaneous. We assume that the transmission of classical or quantum systems between laboratories is free of errors and at the speed of light. Bob has secure and error-free quantum channels between his laboratories. Alice has secure and error-free classical channels between hers. There are not any losses of the transmitted classical or quantum states. Finally, we require that the spacetime regions $R_0$ and $R_1$ at which Bob needs to produce his outputs are secure from Alice.

\section{Unconditionally secure SCOT}
\label{SIV}
We present an unconditionally secure quantum SCOT protocol in Minkowski spacetime. Alice's input $\bold{x}_i=(x_i^{0},x_i^{1},\ldots,x_i^{n-1})\in\{0,1\}^n$ is generated randomly and securely by $\mathcal{A}$ in the past light cone of $P$, and stored securely in the computer $\mathcal{C}_i$ in the past light cone of $Q_i$, for $i=0,1$.

In the past light cone of $P$, $\mathcal{A}$ prepares $\bold{r}=(r_0,r_1,\ldots,r_{n-1})\in\{0,1\}^n$ and $\bold{s}=(s_0,s_1,\ldots,s_{n-1})\in\{0,1\}^n$ randomly, and $n$ qubits $A_0,\ldots, A_{n-1}$ in BB84 states. $A_j$ is prepared in $\lvert \psi_{r_j}^{s_j}\rangle$ for $j=0,1,\ldots,n-1$, where $\lvert \psi_k^0\rangle=\lvert k\rangle$, and $\lvert \psi_k^1\rangle=\frac{1}{\sqrt{2}}\bigl(\lvert 0\rangle+(-1)^{k}\lvert 1\rangle\bigr)$, for $k\in\{0,1\}$, and where $\langle 0\vert 1\rangle=0$. $\mathcal{A}$ gives $\mathcal{B}$ the qubits $A_j$ with their labels $j$ at $P$. Using her secure classical channels, $\mathcal{A}$ sends $\bold{s}$ and $\bold{t}_i=\bold{r}\oplus \bold{x}_i$ to $\mathcal{A}_i$, who receives them at $Q_i$, where $\oplus$ denotes bitwise sum modulo two. $\mathcal{A}_i$ gives $\bold{s}$ and $\bold{t}_i$ to $\mathcal{B}_i$ at $Q_i$.

Immediately after receiving the qubits $A_j$ from $\mathcal{A}$ at $P$, $\mathcal{B}$ sends them through his secure quantum channel to $\mathcal{B}_b$, in a consecutive order according to their labels $j$. At $Q_b$, for each $j=0,1,\ldots,n-1$, $\mathcal{B}_b$ measures $A_j$ in the basis $\mathcal{D}_{s_j}=\{\lvert \psi_{0}^{s_j}\rangle,\lvert \psi_{1}^{s_j}\rangle\}$, obtaining the bit outcome $r_j$. Then, $\mathcal{B}_b$ computes $\bold{r}\oplus \bold{t}_b$, which equals $\bold{x}_b$. Thus, $\mathcal{B}_b$ outputs $\bold{x}_b$ at $Q_b$, as required.

We discuss in Appendix \ref{appb} how this protocol can be modified to allow for a small fraction of errors $\gamma<0.015$. We give in Appendix \ref{appc} an unconditionally secure protocol for SCOT, a slight variation of this protocol, in which the geometry of the output regions $R_i$ are defined differently: the bit $x_i^j$ must be output at the spacetime point $Q_i^j=(h+j\delta,-(-1)^ih,0,0)$ for $j=0,1\ldots,n-1$, where $\delta>0$.

\subsection{Security against Alice}
Our protocol is clearly secure against Alice given our assumptions that Bob's laboratories and quantum channels are secure.

\subsection{Security against Bob}
We note that Bob could easily output $\bold{x}_0$ in $R_0$ and $\bold{x}_1$ in $R_1$, if $R_0$ and $R_1$ were allowed to be timelike separated.
For example, if $R_0$ is defined as above, but $R_1$ is allowed to be the intersection of the future light cone of $\bar{Q}=(2h,0,0,0)$ (see Fig. \ref{fig}) and a small neighborhood of $\bar{Q}$ in the frame $\mathcal{F}$, Bob can play a honest strategy to output $\bold{x}_0$ at $Q_0$ and also output $\bold{x}_1$ at $\bar{Q}$. Agent $\mathcal{B}_0$ obtains $\bold{r}$ at $Q_0$, outputs $\bold{x}_0$ at $Q_0$ and sends $\bold{r}$ to $\mathcal{B}$ at the speed of light. After receiving $\bold{t}_1$ at $Q_1$, $\mathcal{B}_1$ sends $\bold{t}_1$ to $\mathcal{B}$ at the speed of light. Agent $\mathcal{B}$ receives $\bold{r}$ and $\bold{t}_1$ at $\bar{Q}$, where he computes $\bold{r}\oplus \bold{t}_1$, which equals $\bold{x}_1$. Thus, $\mathcal{B}$ outputs $\bold{x}_1$ at $\bar{Q}$. This cheating strategy is consistent with the general Mayers-Lo-Chau quantum attacks \cite{M97,LC97,L97}: Bob can learn $\bold{x}_0$ at $Q_0$ and then implement a nonlocal unitary on the systems held by his agents that allows him to learn also $\bold{x}_1$ at some spacetime point $\bar{Q}$. However, due to Minkowski causality, this nonlocal unitary cannot be completed outside the intersection of the future light cones of $Q_0$ and $Q_1$, hence, $\bar{Q}$ cannot be spacelike separated from $Q_0$ as required by the definition of SCOT. Thus, in general, the spacelike separation of the spacetime points  $Q_0$ and $Q_1$ at which Bob's outputs must be produced guarantees the security of our protocol, as we show below.

Consider an arbitrary cheating strategy by Bob in which he outputs $\bold{x}_0$ at $Q_0$ and $\bold{x}_1$ at $Q_1$ with probability $p_n$. Since in a honest strategy, Bob achieves $p_n=2^{-n}$ by randomly guessing $\bold{x}_{\bar{b}}$, ideally, unconditional security would be defined by satisfaction of a bound $p_n\leq 2^{-n}+f(m)$, where $m$ is a security parameter independent of $n$ such that $f(m) \rightarrow 0$ as $m \rightarrow \infty$. In practice, given the motivation for SCOT in which $\bold{x}_i$ is a password that the computer $\mathcal{C}_i$ accepts as correct if input within $R_i$, we define unconditional security by satisfaction of 
\begin{equation}
\label{eq:a2}
p_n\leq f(m),
\end{equation}
where the security parameter $m$ is allowed to be $m=n$, and where $f(m) \rightarrow 0$ as $m \rightarrow \infty$.
We show below that our SCOT protocol is unconditionally secure according to definition (\ref{eq:a2}) by taking $m=n$. Importantly, we show that $f(n) \rightarrow 0$ exponentially in $n$.

An arbitrary cheating strategy by Bob is as follows. At the spacetime point $P$, agent $\mathcal{B}$ applies a unitary operation $U$ on $A=A_0A_1\cdots A_{n-1}$ and some ancillary system $B$ in a fixed pure state $\lvert \chi\rangle$, then he sends the subsystem $B_0$ of $AB=B_0B_1$ to $\mathcal{B}_0$ and the other subsystem, $B_1$, to $\mathcal{B}_1$, who receive them at the respective spacetime points $Q_0$ and $Q_1$. The Hilbert spaces of $B_0$ and $B_1$ have arbitrary finite dimensions. After receiving $\bold{s}$ and $\bold{t}_i=\bold{x}_i\oplus \bold{r}$ at $Q_i$, $\mathcal{B}_i$ implements a measurement $M_{i,\bold{s}}=\{\Pi_{i,\bold{s}}^{\bold{r}_i}\}_{\bold{r}_i\in\{0,1\}^n}$ on $B_i$ chosen according to the value of $\bold{s}$, obtaining the output $\bold{r}_i=(r_i^0,r_i^1,\ldots,r_i^{n-1})\in\{0,1\}^n$. Finally, at the spacetime point $Q_i$, $\mathcal{B}_i$ outputs $\bold{y}_i=\bold{r}_i\oplus \bold{t}_i$, which equals $\bold{x}_i$ if $\bold{r}_i=\bold{r}$.  

We argue that a general cheating strategy is as described above. Any measurement implemented by $\mathcal{B}$ at $P$ can be delayed until the spacetime points $Q_i$. The measurement implemented by $\mathcal{B}_i$ at $Q_i$ in general depends on $\bold{s}$ but not on $\bold{t}_i=\bold{x}_i\oplus \bold{r}$, because the state $\lvert \psi_{r_j}^{s_j}\rangle$ of qubit $A_j$ depends on $s_j$ but not on $t_i^j$, the $j$th entry of $\bold{t}_i$, which is random. Similarly, in general, the goal of this measurement is to output a $n-$bits string $\bold{r}_i$ which must equal $\bold{r}$ with high probability because of the encoding $\bold{x}_i=\bold{t}_i\oplus \bold{r}$.

We show below that 
\begin{equation}
\label{eq:c1}
p_n\leq\bar{p}_n,
\end{equation}
where $\bar{p}_n\equiv \bigl(\frac{1}{2}+\frac{1}{2\sqrt{2}}\bigr)^n$. Thus, unconditional security, as defined by (\ref{eq:a2}), follows. We note that the bound (\ref{eq:c1}) is achieved by the following simple strategy. Agent $\mathcal{B}$ measures $A_j$ in the Breidbart basis $\{\lvert \omega_0\rangle,\lvert \omega_1\rangle\}$, where $\lvert\omega_0\rangle=\cos(\frac{\pi}{8})\lvert 0\rangle+\sin(\frac{\pi}{8})\lvert 1\rangle$ and $\lvert \omega_1\rangle=-\sin(\frac{\pi}{8})\lvert 0\rangle+\cos(\frac{\pi}{8})\lvert 1\rangle$. Then, $\mathcal{B}$ sends the outcome to $\mathcal{B}_i$, who outputs it as $r_i^j$, for $j=0,1,\ldots,n-1$ and for $i=0,1$. 

We show (\ref{eq:c1}). The proof follows from the result of Ref. \cite{TFKW13}. Tomamichel et al. \cite{TFKW13} consider the following task, which they call $\text{G}_{\text{BB}84}^{\times n}$. Two parties, Bob and Charlie, prepare a tripartite quantum state $\lvert \xi\rangle_{ABC}$. Bob and Charlie keep the respective systems $B$ and $C$, whose Hilbert spaces have arbitrary finite dimensions, and send the system $A=A_0A_1\cdots A_{n-1}$ to another party, Alice, where each system $A_j$ is a qubit. After the preparation of $\lvert\xi\rangle_{ABC}$, Bob and Charlie are not allowed to communicate. Alice randomly generates $\bold{s}=(s_0,\ldots,s_{n-1})\in\{0,1\}^n$ and measures $A_j$ in the BB84 basis  $\mathcal{D}_{s_j}$, obtaining the outcome $r_j\in\{0,1\}$, for $j=0,1,\ldots,n-1$. Alice announces $\bold{s}$ to Bob and Charlie. Bob and Charlie apply measurements $M_{0,\bold{s}}$ and $M_{1,\bold{s}}$, which in general depend on $\bold{s}$, on systems $B$ and $C$, obtaining the $n$-bit outcomes $\bold{r}_0$ and $\bold{r}_1$, respectively. Bob and Charlie win the game if $\bold{r}_0=\bold{r}_1=\bold{r}$, where $\bold{r}=(r_0,r_1,\ldots,r_{n-1})$. It is shown in \cite{TFKW13} that the maximum probability to win the $\text{G}_{\text{BB}84}^{\times n}$ game, optimized over all allowed quantum strategies by Bob and Charlie, is given by $\bar{p}_n$.

We relate a cheating strategy by Bob in our SCOT protocol to a strategy by Bob and Charlie in the $\text{G}_{\text{BB}84}^{\times n}$ game.
In our SCOT protocol, it is equivalent if $\mathcal{A}$ prepares $n$ Bell states $\lvert \Phi^+\rangle_{A'A}=\otimes_{j=0}^{n-1}\lvert \phi^+\rangle_{A_j'A_j}$, she sends $A=A_0A_1\cdots A_{n-1}$ to $\mathcal{B}$ and measures qubit $A_j'$ in the basis $\mathcal{D}_{s_j}$. If Alice obtains outcome $r_j$ for qubit $A_j'$, then qubit $A_j$ projects into the state $\lvert \psi_{r_j}^{s_j}\rangle$. Consider a $\text{G}_{\text{BB}84}^{\times n}$ game in which Bob and Charlie initially prepare $\lvert \xi\rangle_{A'AB}=(\mathds{1}_{A'}\otimes U_{AB})\lvert \Phi^+\rangle_{A'A}\lvert \chi\rangle_B$ and send $A'$ to Alice, Bob keeps $B_0$ and Charlie keeps $B_1$, where $AB=B_0B_1$, and where $U$ and $\lvert\chi\rangle$ are the unitary operation and the state used in the cheating strategy by Bob in our SCOT protocol. Then, by identifying agents $\mathcal{B}_0$ and $\mathcal{B}_1$ in our SCOT protocol with Bob and Charlie in the $\text{G}_{\text{BB}84}^{\times n}$ game, it follows that Bob's cheating probability in our SCOT protocol is upper bounded by $p_n\leq \bar{p}_n$, as claimed.

\section{Discussion}
\label{sec:6}
Here we defined the cryptographic task of spacetime-constrained oblivious transfer (SCOT). We showed that unconditionally secure SCOT cannot be achieved with classical protocols in Minkowski or Galilean spacetime, or with quantum protocols in Galilean spacetime. We gave a quantum SCOT protocol in Minkowski spacetime that we proved unconditionally secure.

It is straightforward to see that our SCOT protocol can be used to implement unconditionally secure bit commitment. Bob commits to a bit $b$ at the spacetime point $P$. Assuming that Alice reads the passwords that Bob inputs into the computers $\mathcal{C}_i$, Bob unveils his commitment by introducing the password $\bold{x}_b$ into $\mathcal{C}_b$ within the required spacetime region $R_b$.

On the other hand, unconditionally secure SCOT cannot in general be obtained from unconditionally secure relativistic bit commitment. This follows because there are unconditionally secure classical relativistic bit commitment protocols \cite{K99,K05.2,LKBHTWZ15,AK15.2}, but unconditionally secure SCOT cannot be achieved with classical protocols, as we proved here. It would be interesting to investigate under what general conditions unconditionally secure SCOT can be used to implement unconditionally secure relativistic bit commitment.

Furthermore, we note that our SCOT protocol seems quite similar to Kent's quantum relativistic bit commitment protocol \cite{K12}, in that in both protocols random BB84 states are transmitted at the spacetime point $P$. An important difference between our SCOT protocol and Kent's bit commitment protocol \cite{K12} is that in our protocol, at the spacetime point $Q_i$, Bob's agent $\mathcal{B}_i$ is informed of the bases on which Alice prepared the transmitted qubits, while in Kent's protocol the committer (who would be Bob in our notation, but is Alice in the notation of Refs. \cite{K12,CK12}) is never revealed the bases by the other party. This implies that the security proofs of Kent's protocol \cite{K12}, for example the one by Croke and Kent \cite{CK12}, cannot in general be adapted to show the security of our SCOT protocol, because in a general cheating strategy by Bob in our protocol, Bob's agents $\mathcal{B}_i$ apply measurements that depend on the bases on which Alice prepared the qubits, which is not possible in the cheating strategies in the bit commitment protocol \cite{K12}.

The defined task of SCOT can be generalized to settings in which Alice has $m$ inputs $\bold{x}_0,\bold{x}_1,\ldots,\bold{x}_{m-1}\in\{0,1,\ldots,d-1\}$ that Bob must output within spacelike separated regions $R_0,R_1,\ldots,R_{m-1}$ according to his input $b\in\{0,1,\ldots,m-1\}$.
One can also investigate different configurations of the geometry for the output and input regions (see an example in Appendix \ref{appc}). Additionally, one can relax the condition that Alice should not be able to learn Bob's input $b$ anywhere in spacetime to the weaker requirement that Alice should not learn $b$ within some fixed particular region of spacetime.

It would be interesting to investigate whether SCOT protocols can be proved unconditionally secure from fundamental physical principles in more general models of spacetime or in postquantum theories of correlations. For example, can we find SCOT protocols that are provably unconditionally secure from the violation of Bell inequalities and the no-signaling principle (as recently done for relativistic quantum bit commitment protocols \cite{AK15.2})?

Finally, it is interesting to investigate whether SCOT can be combined with other cryptographic tasks to perform more general spacetime-constrained multi-party computation and other sophisticated cryptographic tasks.
\begin{acknowledgments}
The author thanks Jonathan Silman, Serge Massar and Stefano Pironio for very helpful conversations. This work was mainly done at the Laboratoire d'Information Quantique, Universit\'e libre de Bruxelles, with financial support from the European Union under the project QALGO, from the F.R.S.-FNRS under the project DIQIP and from the InterUniversity Attraction Poles of the Belgian Federal Government through project Photonics@be. Part of this work was completed at IRIF, Universit\'e Paris Diderot, supported by the project ERC QCC.
\end{acknowledgments}

\appendix
\section{\label{appa}Details of the impossibility proof for unconditionally secure classical protocols for SCOT}

We show (\ref{2}). For fixed given values of $\bold{x}_0,\bold{x}_1$, we define the sets $\bar{\mathcal{G}}^0\equiv \{\bold{g}\in\bar{\mathcal{G}}\vert P(\bold{g}\vert \bold{x}_0,\bold{x}_1,0) \geq P(\bold{g}\vert \bold{x}_0,\bold{x}_1,1)\}$ and $\bar{\mathcal{G}}^1\equiv \{\bold{g}\in\bar{\mathcal{G}}\vert P(\bold{g}\vert \bold{x}_0,\bold{x}_1,1)> P(\bold{g}\vert\bold{x}_0,\bold{x}_1,0)\}$. We consider a cheating strategy by Alice in which she first follows the honest protocol and then guesses that $b=i$ if a communication event $\bold{g}\in \bar{\mathcal{G}}^i$ takes place. Alice's probability to guess $b$ correctly, given her inputs $\bold{x}_0,\bold{x}_1$, is thus 
\begin{equation}
\label{3}
P_{\text{A}}^{\text{c}}=\frac{1}{2}\sum_{i=0}^1 \sum_{\bold{g}\in \bar{\mathcal{G}}^i}P(\bold{g}\vert \bold{x}_0,\bold{x}_1,i).
\end{equation}
 It follows that 
\begin{eqnarray}
\label{4}
P_{\text{A}}^{\text{c}}&=&\frac{1}{2}+\frac{1}{2}\sum_{\bold{g}\in \bar{\mathcal{G}}^i}\Bigl(P\bigl(\bold{g}\vert \bold{x}_0,\bold{x}_1,i\bigr)-P\bigl(\bold{g}\vert \bold{x}_0,\bold{x}_1,\bar{i}\bigr)\Bigr)\nonumber\\
&=&\frac{1}{2}+\frac{1}{2}\sum_{\bold{g}\in \bar{\mathcal{G}}^i}\Bigl\lvert P\bigl(\bold{g}\vert \bold{x}_0,\bold{x}_1,i\bigr)-P\bigl(\bold{g}\vert \bold{x}_0,\bold{x}_1,\bar{i}\bigr)\Bigr\rvert,\nonumber\\
\end{eqnarray}
for $i=0,1$, where in the second line we used that from the definition of $\bar{\mathcal{G}}^i$, we have that $P\bigl(\bold{g}\vert \bold{x}_0,\bold{x}_1,i\bigr)\geq P\bigl(\bold{g}\vert \bold{x}_0,\bold{x}_1,\bar{i}\bigr)$ for all $\bold{g}\in\bar{\mathcal{G}}^i$ and $i\in\{0,1\}$. It follows from (\ref{4}) that 
\begin{equation}
\label{5}
P_{\text{A}}^{\text{c}}= \frac{1}{2}+\frac{1}{4}\sum_{\bold{g}\in \bar{\mathcal{G}}}\bigl\lvert P(\bold{g}\vert \bold{x}_0,\bold{x}_1,0)-P(\bold{g}\vert \bold{x}_0,\bold{x}_1,1)\bigr\rvert.
\end{equation}
Thus, we see from (\ref{5}) that satisfaction of (\ref{1}) implies that (\ref{2}) must hold.

We show (\ref{d1.1}). The probability that Bob outputs $\bold{x}_i$ at $Q_i$ correctly following the honest protocol with input $b=i$ is given by
\begin{equation}
\label{d2}
P_{\text{B}}^{\text{h},i}=\frac{1}{2^{2n}}\sum_{\bold{x}_0,\bold{x}_1}\sum_{\bold{g}\in\bar{\mathcal{G}}} P_{\text{B}}^{\text{h}}(\bold{x}_i\vert \bold{g},\bold{x}_0,\bold{x}_1,i)P(\bold{g}\vert \bold{x}_0,\bold{x}_1,i),
\end{equation}
where $P_{\text{B}}^{\text{h}}(\bold{x}_i\vert \bold{g},\bold{x}_0,\bold{x}_1,i)$ is the probability that Bob guesses the value of $\bold{x}_i$ correctly at the spacetime point $Q_i$ following the honest protocol with input $b=i$ given that Alice's inputs are $\bold{x}_0,\bold{x}_1$ and the communication event $\bold{g}\in\bar{\mathcal{G}}$ takes place, for $i\in\{0,1\}$. From (\ref{d1}), the probability that Bob succeeds in the honest protocol satisfies
\begin{equation}
\label{d3}
P_{\text{B}}^{\text{h},i}\geq 1-\epsilon.
\end{equation}

Consider a cheating strategy by Bob in which he initially sets his input $b=i$ and runs a honest protocol, obtaining communication events $g$, $g_0$ and $g_1$ at $P$, $Q_0$ and $Q_1$, respectively. Bob thus outputs $\bold{x}_i$ at $Q_i$ with probability $P_{\text{B}}^{\text{h},i}$. However, in addition to following the honest protocol with input $b=i$, Bob's agents perform the following actions. They produce perfect copies of the communicated messages $g$, $g_0$ and $g_1$ at $P$, $Q_0$ and $Q_1$, respectively. They use these copies to simultaneously run a honest protocol with the other input $b=\bar{i}$, obtaining the output $\bold{x}_{\bar{i}}$ at $Q_{\bar{i}}$ with some probability $P_{\text{B}}^{\bar{i}}$. The success probability of this simultaneous protocol is given by
\begin{eqnarray}
\label{d4}
P_{\text{B}}^{\bar{i}}&=&\frac{1}{2^{2n}}\sum_{\bold{x}_0,\bold{x}_1}\sum_{\bold{g}\in\bar{\mathcal{G}}} P_{\text{B}}^{\text{h}}\bigl(\bold{x}_{\bar{i}}\vert \bold{g},\bold{x}_0,\bold{x}_1,\bar{i}\bigr)P\bigl(\bold{g}\vert \bold{x}_0,\bold{x}_1,i\bigr)\nonumber\\
&=&P_{\text{B}}^{\text{h},\bar{i}}+\frac{1}{2^{2n}}\sum_{\bold{x}_0,\bold{x}_1}\sum_{\bold{g}\in\bar{\mathcal{G}}} P_{\text{B}}^{\text{h}}\bigl(\bold{x}_{\bar{i}}\vert \bold{g},\bold{x}_0,\bold{x}_1,\bar{i}\bigr)\Delta(\bold{g}, \bold{x}_0,\bold{x}_1,i)\nonumber\\
&\geq&P_{\text{B}}^{\text{h},\bar{i}}-\frac{1}{2^{2n}}\sum_{\bold{x}_0,\bold{x}_1}\sum_{\bold{g}\in\bar{\mathcal{G}}} P_{\text{B}}^{\text{h}}\bigl(\bold{x}_{\bar{i}}\vert \bold{g},\bold{x}_0,\bold{x}_1,\bar{i}\bigr)\bigl\lvert \Delta(\bold{g},\bold{x}_0,\bold{x}_1,i)\bigr\rvert\nonumber\\
&\geq&P_{\text{B}}^{\text{h},\bar{i}}-\frac{1}{2^{2n}}\sum_{\bold{x}_0,\bold{x}_1}\sum_{\bold{g}\in\bar{\mathcal{G}}} \bigl\lvert \Delta(\bold{g},\bold{x}_0,\bold{x}_1,i)\bigr\rvert\nonumber\\
&\geq&P_{\text{B}}^{\text{h},\bar{i}}-4\delta,
\end{eqnarray}
where $\Delta(\bold{g}, \bold{x}_0,\bold{x}_1,i)\equiv P\bigl(\bold{g}\vert \bold{x}_0,\bold{x}_1,i\bigr)-P\bigl(\bold{g}\vert \bold{x}_0,\bold{x}_1,\bar{i}\bigr)$, in the second line we used $P\bigl(\bold{g}\vert \bold{x}_0,\bold{x}_1,i\bigr)=P\bigl(\bold{g}\vert \bold{x}_0,\bold{x}_1,\bar{i}\bigr)+\Delta\bigl(\bold{g}, \bold{x}_0,\bold{x}_1,i\bigr)$ and expression (\ref{d2}), and in the last line we used (\ref{2}). It follows from (\ref{d3}) and (\ref{d4}) that Bob's success probability in this simultaneous protocol satisfies $P_{\text{B}}^{\bar{i}}\geq 1-\epsilon-4\delta$. Since $\epsilon<<1$ and $\delta<<1$, Bob obtains both $\bold{x}_0$ at $Q_0$ and $\bold{x}_1$ at $Q_1$ with a probability $P_{\text{B}}^{\text{c}}$ close to unity: $P_{\text{B}}^{\text{c}}\geq P_{\text{B}}^{\bar{i}}+P_{\text{B}}^{\text{h},i}-1\geq 1-2\epsilon-4\delta$, for any $i\in\{0,1\}$, which shows (\ref{d1.1}).

\section{\label{appb}Allowing for errors}
The SCOT protocol described in the main text can be modified to allow for a small fraction of errors $\gamma<0.015$. A password $\bold{y}_i$ is accepted as valid by $\mathcal{C}_i$ if it differs from $\bold{x}_i$ in at most $\gamma n$ bit entries. Under this condition, the probability $p_n^\gamma$ that both $\mathcal{B}_0$ and $\mathcal{B}_1$ input valid passwords into $\mathcal{C}_0$ and $\mathcal{C}_1$, respectively, is bounded by $p_n^\gamma \leq \bigl(2^{2h(\gamma)}\bar{p}_1\bigr)^n$, where $h(\gamma)=-\gamma\log_2 \gamma-(1-\gamma)\log_2 (1-\gamma)$ is the binary entropy. This bound follows because the probability to win a modified version of the $\text{G}_{\text{BB}84}^{\times n}$ in which a fraction $\gamma$ of bit errors is allowed is not greater than $\bigl(2^{2h(\gamma)}\bar{p}_1\bigr)^n$ \cite{TFKW13}. Thus, for $\gamma<0.015$, we have $2^{2h(\gamma)}\bar{p}_{1}<1$, hence, unconditional security follows.

\section{\label{appc}SCOT with modified geometry of the output regions}
We consider a SCOT scenario in which the geometry of the output regions $R_i$ is defined as follows. The bit $x_i^j$ must be output at the spacetime point $Q_i^j=(h+j\delta,-(-1)^ih,0,0)$ for $j=0,1\ldots,n-1$, where $\delta>0$. We say that Bob outputs $\bold{x}_i$ within $R_i$ if he outputs $x_i^j$ at $Q_i^j$ for all $j=0,1,\ldots,n-1$.

We give an unconditionally secure SCOT protocol for the output regions defined above. In the past light cone of $P$, $\mathcal{A}$ prepares $\bold{r}=(r_0,r_1,\ldots,r_{n-1})\in\{0,1\}^n$ and $\bold{s}=(s_0,s_1,\ldots,s_{n-1})\in\{0,1\}^n$ randomly, and $n$ qubits $A_0,\ldots, A_{n-1}$ in BB84 states. The qubit $A_j$ is prepared in the state $\lvert \psi_{r_j}^{s_j}\rangle$. $\mathcal{A}$ gives $\mathcal{B}$ the qubit 
$A_j$ at the spacetime point $P_j=(j\delta,0,0,0)$. Using her secure classical channels, $\mathcal{A}$ sends the bits $s_j$ and $t_i^j=r_j\oplus x_i^j$ to $\mathcal{A}_i$, who receives them at $Q_i^j$. Agent $\mathcal{A}_i$ gives agent $\mathcal{B}_i$ these bits at $Q_i^j$.

Immediately after receiving $A_j$ from $\mathcal{A}$ at $P_j$, $\mathcal{B}$ sends it through his secure quantum channel to $\mathcal{B}_b$. At $Q_b^j$, $\mathcal{B}_b$ measures $A_j$ in the BB84 basis $\mathcal{D}_{s_j}$, obtaining the bit outcome $r_j$. Then, $\mathcal{B}_b$ computes the bit $r_j\oplus t_b^j$, which equals $x_b^j$. Thus, $\mathcal{B}_b$ outputs $x_b^j$ at $Q_b^j$, for all $j=0,1,\ldots,n-1$, as required.

\subsection{Security against Alice}
Similarly to the SCOT protocol given in the main text, this protocol is secure against Alice, given our assumptions that Bob's laboratories and quantum channels are secure.

\subsection{Security against Bob}
The security against Bob follows trivially from the security proof of the SCOT protocol given in the main text. We give an alternative security proof below.

In an arbitrary cheating strategy by Bob, at the spacetime point $P_j$, agent $\mathcal{B}$ applies a unitary operation $U_j$ on the qubit $A_j$, received from $\mathcal{A}$, and some ancillary system $E_j$ in a fixed pure state $\lvert \chi\rangle$, then he sends the subsystem $B_0^j$ of $A_jE_j=B_0^jB_1^j$ to his agent $\mathcal{B}_0$ and the other subsystem, $B_1^j$, to his agent $\mathcal{B}_1$, who receive them at the respective spacetime points $Q_0^j$ and $Q_1^j$. After receiving the bits $s_j$ and $t_i^j=x_i^j\oplus r_j$ at $Q_i^j$, $\mathcal{B}_i$ implements a measurement on $B_i^j$ chosen according to the value of $s_j$, obtaining a bit output $r_i^j$. Finally, at the spacetime point $Q_i^j$, $\mathcal{B}_i$ outputs $y_i^j=r_i^j\oplus t_i^j$ , which equals $x_i^j$ if $r_i^j=r_j$.  

We argue that a general cheating strategy is of the form described above. First, since the spacetime point $Q_i^j$ where $\mathcal{B}_i$ must output $x_i^j$ is light like separated from the spacetime point $P_j$ where agent $\mathcal{B}$ receives the qubit $A_j$ encoding $x_i^j$, $\mathcal{B}$ must apply any quantum operations on $A_jE_j$ at $P_j$ and instantaneously send $B_i^j$ to $\mathcal{B}_i$. This implies that $\mathcal{B}$ cannot wait a time $\delta$ to receive the next qubit $A_{j+1}$ at $P_{j+1}$ in order to apply a joint operation on $A_j$ and $A_{j+1}$.

Second, an arbitrary measurement that $\mathcal{B}$ might perform at $P_j$ can be delayed until points $Q_i^j$. In particular, $\mathcal{B}$ cannot obtain any information about the bits $r_{j'}$ with $j'>j$ encoded in subsequent qubits $A_{j+1},A_{j+2},\ldots,A_{n-1}$ by applying a measurement on $A_j$ because the bits $r_j$ and $s_j$ are random, hence, the state $\lvert \psi_{r_j}^{s_j}\rangle$ of qubit $A_j$ does not encode any information about bits $r_{j'}$ or $x_i^{j'}$ for $j'\neq j$. Similarly, $\mathcal{B}$ cannot obtain any information about $r_{j'}$ that is not already encoded in $A_{j'}$ by preparing some ancilla $D_j$ entangled with $A_j$ and then applying some joint measurement on $D_jA_{j'}$, for $j'>j$.

Finally, the measurement implemented by $\mathcal{B}_i$ at $Q_i^j$ in general depends on $s_j$ but not on $t_i^j=x_i^j\oplus r_j$ because the state $\lvert \psi_{r_j}^{s_j}\rangle$ of qubit $A_j$ depends on $s_j$ but not on $t_i^j$, which is random. Similarly, in general, the goal of this measurement is to output a bit $r_i^j$ that equals $r_j$ with high probability because of the encoding of $x_i^j$ via $t_i^j=x_i^j\oplus r_j$.

Let $q^j$ be the maximum success probability that $\mathcal{B}_0$ and $\mathcal{B}_1$ output $x_0^j$ and $x_i^j$ at $Q_0^j$ and $Q_1^j$, respectively. 
As discussed above, in general, Bob must treat qubit $A_j$ independently of other qubits $A_{j'}$. Since the protocol that Alice follows to encode $x_i^j$ is the same for any $j$, and bits $x_i^j$ and $x_i^{j'}$ are encoded independently for any $j'\neq j$, we have that $q^j=q^{j'}\equiv q$ for all $j,j'\in\{0,1\ldots,n-1\}$, which is achieved by the same strategy on $A_jE_j$ for all $j$. We show below that 
\begin{equation}
\label{eq:a2.1}
q\leq \frac{1}{2}\Bigl(1+\frac{1}{\sqrt{2}}\Bigr).
\end{equation}
Since the probability that $\mathcal{B}_i$ outputs $x_i^j$ at $Q_i^j$ is independent of the probability to output $x_{i'}^{j'}$ for $j'\neq j$, as follows from the discussion above, we have that the maximum success probability that $\mathcal{B}_0$ and $\mathcal{B}_1$ output $x_0^j$ and $x_1^j$ at $Q_0^j$ and $Q_1^j$, respectively, for all $j=0,1,\ldots,n-1$ is $p_n^{\text{max}}=(q)^n$. It follows from (\ref{eq:a2.1}) that unconditional security, as defined by equation (1) of the main text, is satisfied.

We show (\ref{eq:a2.1}) by considering the case $j=0$. We simplify notation by setting $A_0=A$, $E_0=E$, $B_i^0=B_i$, $P_0=P$, $Q_i^0=Q_i$,  $x_i^0=x_i$, $r_0=r$, $s_0=s$ and $r_i^0=r_i$. Let $q_i$ be the probability that $\mathcal{B}_i$ outputs $x_i$ at $Q_i$, for $i\in\{0,1\}$, in a strategy achieving $q$. It follows that $q\leq \frac{1}{2}(q_0+q_1)$. We show below that 
\begin{equation}
\label{eq:a2.2}
q_{\text{max}}=1+\frac{1}{\sqrt{2}},
\end{equation}
where $q_{\text{max}}\equiv \max\{q_0+q_1\}$, and where the maximum is taken over all possible strategies by Bob, as described above. Thus, (\ref{eq:a2.1}) follows. We note that $q_{\text{max}}$ is achieved if Bob applies $1\rightarrow 2$ optimal symmetric cloning \cite{SIGA05} of the BB84 states to qubit $A$ and sends a clone to $\mathcal{B}_i$, who then measures it in the BB84 basis $\mathcal{D}_{s}$ on which $A$ was prepared and outputs the outcome as $r_i$, for $i\in\{0,1\}$. Alternatively, $q_{\text{max}}$ is achieved if $\mathcal{B}$ measures $A$ in the Breidbart basis $\{\lvert \omega_0\rangle,\lvert \omega_1\rangle\}$, where $\lvert\omega_0\rangle=\cos(\frac{\pi}{8})\lvert 0\rangle+\sin(\frac{\pi}{8})\lvert 1\rangle$ and $\lvert \omega_1\rangle=-\sin(\frac{\pi}{8})\lvert 0\rangle+\cos(\frac{\pi}{8})\lvert 1\rangle$ and sends the outcome to agents $\mathcal{B}_i$, who outputs it as $r_i$.

We show (\ref{eq:a2.2}). After $\mathcal{B}$ applies the unitary operation $U$, the system $AE=B_0B_1$ is transformed into
\begin{equation}
\label{eq:a3}
\lvert \Psi_{r}^{s}\rangle_{B_0B_1}=U\lvert\psi_{r}^{s}\rangle_A\lvert\chi\rangle_{E}.
\end{equation}
In general, we can consider that the measurement applied by $\mathcal{B}_i$ on $B_i$ is a projective measurement $M_{i,s}=\{\Pi_{i,s}^k\}_{k=0}^1$. This is because an arbitrary measurement can be implemented by introducing an ancilla, included in $B_i$ at $P$, and then performing a projective measurement on the whole system. Thus, we have
\begin{equation}
\label{eq:a4}
q_0+q_1=\frac{1}{4}\sum_{r=0}^1\sum_{s=0}^1\Bigl(\langle \Psi_{r}^{s}\rvert \mathds{1}\otimes\Pi_{1,s}^r\rvert \Psi_{r}^{s}\rangle
+\langle \Psi_{r}^{s}\rvert \Pi_{0,s}^r\otimes \mathds{1}\rvert \Psi_{r}^{s}\rangle\Bigr),
\end{equation}
where the projector $\Pi_{i,s}^r$ acts on system $B_i$ and $\mathds{1}$ is the identity operator acting on the other system. We write
$4(q_0+q_1)=\mu_0+\mu_1$, where
\begin{eqnarray}
\label{eq:a5.1}
\mu_0&=&\langle \Psi_{0}^{0}\rvert \Pi_{0,0}^0\otimes \mathds{1}\rvert \Psi_{0}^{0}\rangle
+\langle \Psi_{1}^{0}\rvert \Pi_{0,0}^1\otimes \mathds{1}\rvert \Psi_{1}^{0}\rangle\nonumber\\
&&\quad+\langle\Psi_{0}^{1}\rvert  \mathds{1} \otimes\Pi_{1,1}^0\rvert \Psi_{0}^{1}\rangle+\langle\Psi_{1}^{1}\rvert  \mathds{1} \otimes\Pi_{1,1}^1\rvert \Psi_{1}^{1}\rangle,\\
\label{eq:a5.2}
\mu_1&=&\langle \Psi_{0}^{0}\rvert \mathds{1}\otimes \Pi_{1,0}^0\rvert \Psi_{0}^{0}\rangle
+\langle \Psi_{1}^{0}\rvert  \mathds{1}\otimes\Pi_{1,0}^1\rvert \Psi_{1}^{0}\rangle
\nonumber\\
&&\quad+\langle\Psi_{0}^{1}\rvert \Pi_{0,1}^0 \otimes \mathds{1} \rvert\Psi_{0}^{1}\rangle+\langle\Psi_{1}^{1}\rvert \Pi_{0,1}^1\otimes  \mathds{1} \rvert \Psi_{1}^{1}\rangle.
\end{eqnarray}
We note from the symmetry of the terms $\mu_i$ that the maximum value of $\mu_0$, $\mu_0^\text{max}$, equals the maximum value of $\mu_1$. Thus, $q_0+q_1\leq \mu_0^\text{max}/2$. We show below that
\begin{equation}
\label{eq:a6}
\mu_0^\text{max}=2+\sqrt{2}.
\end{equation}
Since, as mentioned above, the value $q_0+q_1=1+\frac{1}{\sqrt{2}}$ is achieved by an optimal cloning strategy, or by measuring in the Breidbart basis, equation (\ref{eq:a2.2}) follows.

We show (\ref{eq:a6}). We recall the definitions given in the main text:
\begin{equation}
\label{eq:a1}
\lvert \psi_k^0\rangle=\lvert k\rangle,\qquad \lvert \psi_k^1\rangle=\frac{1}{\sqrt{2}}\bigl(\lvert 0\rangle+(-1)^{k}\lvert 1\rangle\bigr),
\end{equation}
for $k\in\{0,1\}$, and where $\langle 0\vert 1\rangle=0$. From (\ref{eq:a3}) and (\ref{eq:a1}), we have
\begin{equation}
\label{eq:a7}
\langle \Psi_0^0 \vert \Psi_1^0\rangle=0,\qquad \lvert \Psi_r^1\rangle=\frac{1}{\sqrt{2}}\bigl(\lvert \Psi_0^0\rangle+(-1)^r\lvert \Psi_1^0 \rangle\bigr).
\end{equation}
Using the second equality of (\ref{eq:a7}) and the relations $\bigl(\Pi_{i,s}^r\bigr)^\dagger=\Pi_{i,s}^r$ and $\Pi_{i,s}^1=\mathds{1}_{B_i}-\Pi_{i,s}^0$ in (\ref{eq:a5.1}), we reduce the expression of $\mu_0$ to
\begin{eqnarray}
\label{eq:a8}
\mu_0&=&2+\langle \Psi_0^0 \rvert \Pi_{0,0}^0\otimes\mathds{1}\lvert\Psi_0^0\rangle-\langle \Psi_1^0 \rvert \Pi_{0,0}^0\otimes\mathds{1}\lvert\Psi_1^0\rangle\nonumber\\
&&\quad+2\text{Re}\bigl(\langle \Psi_0^0 \rvert \mathds{1}\otimes \Pi_{1,1}^0\lvert\Psi_1^0\rangle\bigr).
\end{eqnarray} 
We express $\lvert \Psi_j^0\rangle$ in term of states $\lvert \alpha_{jkl}\rangle$ living in the subspaces on which
the projectors $\Pi_{0,0}^k$ and $\Pi_{1,1}^l$ act, for $j,k,l\in\{0,1\}$. We have
\begin{equation}
\label{eq:a9}
\lvert \Psi_j^0\rangle=\sum_{k=0}^1 \sum_{l=0}^1 a_{jkl}\lvert \alpha_{jkl}\rangle,
\end{equation}
where we impose the normalization conditions $\langle\alpha_{jkl}\vert \alpha_{jkl}\rangle=1$ and
\begin{equation}
\label{eq:a10}
\sum_{k=0}^1\sum_{l=0}^1\lvert a_{jkl}\rvert^2=1,
\end{equation}
and where the following relation holds:
\begin{equation}
\label{eq:a11}
\bigl(\Pi_{0,0}^k\otimes\Pi_{1,1}^l\bigr)\lvert \alpha_{jkl}\rangle=\lvert \alpha_{jkl}\rangle.
\end{equation}
Using the properties of projectors, $\Pi_{i,s}^{r}\Pi_{i,s}^{r'}=\delta_{r,r'}\Pi_{i,s}^{r}$ and $\bigl(\Pi_{i,s}^{r}\bigr)^\dagger=\Pi_{i,s}^{r}$, we obtain from (\ref{eq:a8}), (\ref{eq:a9}) and (\ref{eq:a11}) that
\begin{eqnarray}
\label{eq:a12}
\mu_0&=&2+2\text{Re}\bigl(a_{000}^*a_{100}\langle \alpha_{000} \vert\alpha_{100}\rangle+a_{010}^*a_{110}\langle \alpha_{010}\vert\alpha_{110}\rangle\bigr)\nonumber\\
&&\qquad+\lvert a_{000}\rvert^2+\lvert a_{001}\rvert^2-\lvert a_{100}\rvert^2-\lvert a_{101}\rvert^2.
\end{eqnarray} 
From (\ref{eq:a9}), (\ref{eq:a11}) and the fact that $\Pi_{i,s}^r$ are projectors, the orthogonality constraint in (\ref{eq:a7}) reads
\begin{equation}
\label{eq:a13}
\langle \Psi_0^0 \vert \Psi_1^0\rangle=\sum_{k=0}^1\sum_{l=0}^1 a_{0kl}^*a_{1kl}\langle \alpha_{0kl}\vert \alpha_{1kl}\rangle=0.
\end{equation}
Using the expression (\ref{eq:a12}), subject to constraints (\ref{eq:a10}), (\ref{eq:a13}) and $\lvert\langle \alpha_{0kl}\vert \alpha_{1kl}\rangle\rvert\leq 1$, we obtained numerically that the maximum of $\mu_0$ is given by (\ref{eq:a6}), as claimed.

We give details of the numerical computation of $\mu_0^\text{max}$, which we performed with the Mathematica software.
We give below the mathematical expression input to the Mathematica program.

The complex variables $a_{jkl}$ and $\langle \alpha_{0kl}\vert \alpha_{1kl}\rangle$ are defined in terms of real variables $c_{jkl}$, $d_{jkl}$, $g_{kl}$ and $h_{kl}$ for $j,k,l\in\{0,1\}$, as follows:
\begin{eqnarray}
\label{b1}
a_{jkl}&=&c_{jkl}+id_{jkl},\\
\langle \alpha_{0kl}\vert \alpha_{1kl}\rangle&=&g_{kl}+ih_{kl}.
\end{eqnarray}
The constraints given by equations (\ref{eq:a10}) and (\ref{eq:a13}) respectively read
\begin{equation}
\label{b2}
\sum_{k=0}^1\sum_{l=0}^1\Bigl((c_{jkl})^2+(d_{jkl})^2\Bigr)=1,
\end{equation}
and
\begin{eqnarray}
\label{b3}
0&=&g_{00}(c_{000}c_{100}+d_{000}d_{100})-h_{00}(c_{000}d_{100}-c_{100}d_{000})\nonumber\\
&&\!\!\!\!\!\!\!\!\quad+g_{01}(c_{001}c_{101}+d_{001}d_{101})-h_{01}(c_{001}d_{101}-c_{101}d_{001})\nonumber\\
&&\!\!\!\!\!\!\!\!\quad+g_{10}(c_{010}c_{110}+d_{010}d_{110})-h_{10}(c_{010}d_{110}-c_{110}d_{010})\nonumber\\
&&\!\!\!\!\!\!\!\!\quad+g_{11}(c_{011}c_{111}+d_{011}d_{111})-h_{11}(c_{011}d_{111}-c_{111}d_{011}),\nonumber\\
0&=&g_{00}(c_{000}d_{100}-c_{100}d_{000})+h_{00}(c_{000}c_{100}+d_{000}d_{100})\nonumber\\
&&\!\!\!\!\!\!\!\!\quad+g_{01}(c_{001}d_{101}-c_{101}d_{001})+h_{01}(c_{001}c_{101}+d_{001}d_{101})\nonumber\\
&&\!\!\!\!\!\!\!\!\quad+g_{10}(c_{010}d_{110}-c_{110}d_{010})+h_{10}(c_{010}c_{110}+d_{010}d_{110})\nonumber\\
&&\!\!\!\!\!\!\!\!\quad+g_{11}(c_{011}d_{111}-c_{111}d_{011})+h_{11}(c_{011}c_{111}+d_{011}d_{111}).\nonumber\\
\end{eqnarray}
The constraints $\lvert\langle \alpha_{0kl}\vert \alpha_{1kl}\rangle\rvert\leq 1$ are given by
\begin{equation}
\label{b4}
(g_{kl})^2+(h_{kl})^2\leq 1. 
\end{equation}
The expression for $\mu_0$ given by equation (\ref{eq:a12}) corresponds to
\begin{eqnarray}
\label{b5}
\mu_0&=&2+\sum_{l=0}^1\Bigl((c_{00l})^2+(d_{00l})^2-(c_{10l})^2-(d_{10l})^2\Bigr)\nonumber\\
&&\quad+2\Bigl(g_{00}(c_{000}c_{100}+d_{000}d_{100})\nonumber\\
&&\qquad\qquad+h_{00}(c_{100}d_{000}-c_{000}d_{100})\nonumber\\
&&\qquad\qquad+g_{10}(c_{010}c_{110}+d_{010}d_{110})\nonumber\\
&&\qquad\qquad+h_{10}(c_{110}d_{010}-c_{010}d_{110})\Bigr).
\end{eqnarray}

We computed the maximum value of $\mu_0$ given by (\ref{b5}), subject to the constraints (\ref{b2}) -- (\ref{b4}).
It was found that $\mu_0^\text{max}=2+\sqrt{2}$, as given by equation (\ref{eq:a6}).

%

\end{document}